\documentclass[prl,aps,twocolumn,preprintnumbers, showpacs, nofootinbib,superscriptaddress,notitlepage]{revtex4-1}
\usepackage{mathrsfs}
\usepackage{amsfonts}
\usepackage{amsmath}
\usepackage{slashed}
\usepackage{array}
\usepackage{verbatim}
\usepackage{epsfig}
\usepackage{graphicx}
\usepackage{color}
\usepackage{hyperref}
\usepackage{footmisc}

\usepackage[dvipsnames]{xcolor}
\usepackage{subfigure}
\usepackage{float}
\usepackage{amssymb}
\usepackage{pifont}
\usepackage{setspace}
\usepackage{multirow}
\usepackage{booktabs}
\usepackage[normalem]{ulem}

\newcommand{\beq}{\begin{eqnarray}}
\newcommand{\eeq}{\end{eqnarray}}

\begin{document}
\title{Proton's isovector PDF with updated analysis of large-momentum lattice data}

\author{Xiangdong Ji}
\affiliation{Tsung-Dao Lee Institute and School of Physics and Astronomy, Shanghai Jiao Tong University, Shanghai 201210, China}

\author{Yushan Su}
\email{Corresponding author: ysu12345@umd.edu}
\affiliation{Department of Physics, University of Maryland, College Park, MD 20742, USA}

\begin{abstract}
The proton's unpolarized $u(x)-d(x)$ parton distribution function (PDF) has been studied by a number of lattice QCD groups through large momentum expansion. However, due to lattice artifacts (excited state contaminations, unphysical pion masses, and discretization effects) and less-advanced theoretical analysis (renormalizations, large-distance extrapolations, and large-log resummations), the resulting PDFs cannot be compared strictly with experimental data. By using the state-of-the-art theoretical tools and mitigating the lattice artifacts empirically, we reanalyze the available datasets in the literature and find that the new PDF for the physical-limit-extrapolated dataset (MSULat20) is consistent with global fits within $\sim1\sigma$, while others are broadly compatible. This provides evidence that large momentum expansion is capable of accurately predicting the $x$-dependence of the PDFs when ideal lattice data become available. 
\end{abstract}

\maketitle

\textit{Introduction:}
\quad The proton's unpolarized isovector parton distribution function (PDF), $u(x)-d(x)$, is among the best-determined from more than 50 years of high-energy experimental data, with uncertainties at a few percent level in the moderate-$x$ region~\cite{ATLAS:2021vod,NNPDF:2021njg,Bailey:2020ooq,Hou:2019efy,Alekhin:2017kpj,H1:2015ubc,Jimenez-Delgado:2014twa}. In recent years, this quantity has been explored from first principles lattice quantum chromodynamics (QCD)~\cite{Wilson:1974sk} within the framework of large momentum expansion (LaMET)~\cite{Ji:2013dva,Ji:2014gla,Ji:2020byp,Ji:2022ezo,Ji:2024oka}, which supposedly has the prediction power of the $x$-dependence. 
The isovector PDF is significantly easier to compute on lattice than the flavor-singlet ones because it avoids ``quark-line-disconnected diagrams''. Therefore, it is an ideal  benchmark observable for the quantitative performance of lattice QCD and large momentum expansion. 

A number of lattice QCD groups have published results on $u(x)-d(x)$ using different lattice configurations with assorted lattice spacings and pion masses~\cite{Lin:2014zya,Alexandrou:2015rja,Chen:2016utp,Alexandrou:2016jqi,Alexandrou:2018pbm,Chen:2018xof,LatticeParton:2018gjr,Alexandrou:2019lfo,Joo:2020spy,Fan:2020nzz,Alexandrou:2020qtt,Lin:2020fsj,Gao:2022uhg,Chen:2024rgi,Gao:2026hix}. After
performing LaMET expansion, the resulting $x$ dependences exhibit noticeable variations and mostly differ from phenomenological determinations. These discrepancies may originate from both lattice artifacts and less-sophisticated theoretical analysis. On the lattice side, excited-state contaminations, unphysical pion masses, and discretization effects remain important sources of systematic uncertainties. Compared with the first-moment calculations~\cite{Mondal:2020ela,Djukanovic:2024krw} where the lattice artifacts have important impacts on the final results, these effects in PDFs are expected to be strong. In fact, they could be even stronger because large momenta result in smaller energy gaps between the ground and excited states and additional suppressions of the signal-to-background ratios. On the theoretical side, 
much theoretical development for large momentum expansion has become available only in recent years~\cite{Ji:2020brr,LatticePartonLPC:2021gpi,Chou:2022drv,Zhang:2023bxs,Ji:2026vir,Li:2020xml,Chen:2020ody,Su:2022fiu,Ji:2023pba,Liu:2023onm,Ji:2024hit}. Most existing calculations predate these developments and therefore have potential systematic errors from incomplete theoretical analysis. 

In this work, we reanalyze the available $u-d$ PDF lattice datasets in Refs.~\cite{LatticeParton:2018gjr,Alexandrou:2020qtt,Lin:2020fsj,Chen:2024rgi} using updated theoretical tools, including hybrid and self-renormalization~\cite{Ji:2020brr,LatticePartonLPC:2021gpi,Chou:2022drv}, large distance extrapolation~\cite{Ji:2026vir}, next-to-next-to-leading order (NNLO) matching~\cite{Li:2020xml,Chen:2020ody}, and perturbative large-log resummations~\cite{Zhang:2023bxs,Su:2022fiu,Ji:2023pba,Liu:2023onm,Ji:2024hit}. To mitigate the majority of lattice artifacts, we subject the data to the constraint of the first moment $\langle x \rangle$ through an empirical procedure. This is 
because the first moment is very sensitive to various lattice artifacts
and has been very challenging to obtain accurately in moment calculations~\cite{Mondal:2020ela,Djukanovic:2024krw}. Therefore, a fair test
of the large-momentum expansion at present is made by assuming that the first moment can already be reproduced in a standard lattice calculation. 
We compare the resulting PDFs with one another and with global fits~\cite{Harland-Lang:2014zoa,Hou:2019efy}, and 
find that the new results are largely consistent with them in physical limits. This provides evidence that large momentum expansion can potentially predict the $x$-dependence of the PDFs when ideal lattice data become available. 

\begin{table*}[htbp]
    \centering
    \begin{tabular}{|c|c|c|c|c|}
      \hline
      Works & MSULat20~\cite{Lin:2020fsj} & ETMC20~\cite{Alexandrou:2020qtt} & LPC18~\cite{LatticeParton:2018gjr} &  CLQCD24~\cite{Chen:2024rgi} \\
      \hline
      $a$ (fm) & 0.12, 0.09, 0.06 &  0.0820 & 0.0854 & 0.105 \\
      \hline
      $m_{\pi}$ (MeV) & $310, 230, 138$ & $373$ & $356$ & $293$ \\
      \hline
      $P^z$ (GeV) & 2.2 & 1.89 & 2.27 & 1.85 \\ 
      \hline
    \end{tabular}
    \caption{Selected lattice QCD datasets for the unpolarized $u-d$ proton quasi-PDF matrix elements. Their information, including lattice spacing $a$, pion mass $m_{\pi}$, and hadron momentum $P^z$, is listed in the table. }
    \label{tab:setlatdat}
\end{table*}

\textit{Lattice QCD datasets:}\quad 
The starting point of our analysis is the $u-d$ unpolarized quasi-PDF matrix elements in a large-momentum proton,
\begin{align}\label{eq:BareM}
\tilde{h}^{B}(z,P^z,a) = 
&\left\langle N(P) \left| \, \left[ \bar{u}(z) U(z,0) \gamma^t u(0) \right. \right. \right. \nonumber\\ 
&\left. \left. \left. - \bar{d}(z) U(z,0) \gamma^t d(0) \right] \, \right| N(P) \right\rangle_B \ ,
\end{align}
where various notations are standard ($|N(P)\rangle$ is a nucleon state of momentum $P^\mu$,  $u$ and $d$ are quark fields, $\gamma^\mu$ the Dirac matrices, $U$ is the Wilson line, and $x^\mu=(t,x,y,z)$ are spacetime coordinates) and ``$B$" denotes the bare matrix elements before renormalization. Previous calculations of this quantity used for large momentum expansion analysis include Refs.~\cite{Lin:2014zya,Alexandrou:2015rja,Chen:2016utp,Alexandrou:2016jqi,Alexandrou:2018pbm,Chen:2018xof,LatticeParton:2018gjr,Alexandrou:2019lfo,Fan:2020nzz,Alexandrou:2020qtt,Lin:2020fsj,Gao:2022uhg,Chen:2024rgi}. 

The excited-state controls in the above works are significantly less sophisticated than the first-moment calculation in Ref.~\cite{Mondal:2020ela}, which demonstrates the level of difficulty in the nucleon matrix elements.  Moreover, only one existing work~\cite{Lin:2020fsj} performs both physical pion mass and continuum extrapolations for the isovector PDF. Two papers~\cite{Gao:2022uhg,Chen:2024rgi} utilize the more advanced hybrid renormalization technique~\cite{Ji:2020brr}. All calculations precede the recent theoretical developments regarding the latest long-distance expansion formulas for spatial correlations~\cite{Ji:2026vir} and complete resummation formulas~\cite{Su:2022fiu,Zhang:2023bxs,Ji:2023pba,Liu:2023onm,Ji:2024hit}. Therefore, these calculations predicted various different $x$-dependences, and many of them deviate from the experimental fittings.

To test if large-momentum expansion can realize the promise 
of predicting the $x$-dependence of PDFs,  we present a new analysis of the lattice data with a better control of the above effects. Considering large momentum (at least $\sim 2$ GeV, corresponding to $v\sim0.9 c$), statistical precision, and data accessibility, we select four characteristic datasets in Refs.~\cite{Lin:2020fsj,Alexandrou:2020qtt,LatticeParton:2018gjr,Chen:2024rgi} for the present analysis, as summarized in Tab.~\ref{tab:setlatdat}. In particular, the MSULat20 dataset~\cite{Lin:2020fsj} allows extrapolations to the continuum and physical pion mass limits, with the coordinate and momentum space statistical uncertainties extracted from Figs.~4 and~5 of the reference, respectively. The other datasets~\cite{Alexandrou:2020qtt,LatticeParton:2018gjr,Chen:2024rgi} are bootstrap samples obtained through private communications. These matrix elements are normalized with respect to the real part at $z=0$ for charge conservation, while the imaginary part at $z=0$ is set to zero according to the hermiticity of the correlator. 

In addition to the construction in Eq.~(\ref{eq:BareM}), the Coulomb gauge method is under exploration~\cite{Gao:2026hix}, and the results are consistent with the global fit PDFs. However, this method requires further theoretical developments, such as next-to-next-to-leading-order matching and threshold log resummation. Therefore, it is not incorporated into this paper's analysis framework. 


\textit{New analysis:}\quad The bare matrix elements in Eq.~(\ref{eq:BareM}) contain the linear and logarithmic divergences~\cite{Alexandrou:2017huk,Ji:2015jwa, Ji:2017oey,Ishikawa:2017faj,Green:2017xeu,Li:2018tpe} regularized by finite lattice spacing $a$, which are absent in renormalized theory. To eliminate these ultraviolet (UV) divergences, early papers proposed or adopted various  approaches~\cite{Green:2017xeu,Chen:2017mzz,Alexandrou:2017huk,Izubuchi:2018srq,Radyushkin:2018cvn,Braun:2018brg}, which, however, suffer from additional infrared (IR) effects at large $z$ that are not perturbatively controllable~\cite{Ji:2020brr,LatticePartonLPC:2021gpi}. To avoid this issue, the hybrid renormalization~\cite{Ji:2020brr,Chou:2022drv} implements an almost minimal UV subtraction while mitigating short-distance discrepancies between lattice and continuum theories. The renormalized matrix elements take the form,
\begin{align}\label{eq:HybridM}
\tilde{h}^{R}(z,P^z) & = \frac{\tilde{h}^{B}(z,P^z,a)}{\tilde{h}^{B}(z,0,a)} \theta\left[ z_s - \left|z\right| \right] \nonumber\\
&+ \frac{\tilde{h}^{B}(z,P^z,a)}{Z_{R}(z,a,\mu)} \frac{Z_{R}(z_s,a,\mu)}{\tilde{h}^{B}(z_s,0,a)} \theta\left[ \left|z\right| - z_s \right] \ .
\end{align}
The cutoff $z_s$ is chosen within the window of $a < z_s \ll 1/\Lambda_{\rm QCD}$. For $|z|<z_s$, the hybrid scheme reduces to the ratio method~\cite{Radyushkin:2017cyf,Orginos:2017kos,Izubuchi:2018srq}, which removes the UV divergences together with part of the discretization effects. For $|z|>z_s$, it is the same as the $\overline{\rm MS}$ scheme up to an overall constant, which subtracts only the UV divergences without introducing extra non-perturbative effects. The factor $Z_{R}(z,a,\mu)$, converting lattice data to the $\overline{\rm MS}$ scheme, is extracted using self renormalization~\cite{LatticePartonLPC:2021gpi}. In particular, $Z_R$ contains a finite renormalization  factor $\sim e^{-m_0 |z|}$, with a new parameter $m_0$. This additional scheme dependence shall be matched, at small $z$, to an infrared-renormalon resummation scheme of the divergent perturbative QCD series in $\alpha_s(z)$, which has exactly the same uncertainty of order ${\cal O}(z\Lambda_{\rm QCD})$~\cite{Pineda:2001zq,Braun:2018brg,Zhang:2023bxs}. More explicitly, we fix the $m_0$ by fitting the renormalized lattice zero momentum matrix element $\tilde{h}^{B}(z,0,a)/Z_{R}(z,a,\mu)$ with the perturbative coefficient $C_0\left(z^2\mu^2\right)$~\cite{Izubuchi:2018srq} at next-to-next-to leading order (NNLO)~\cite{Li:2020xml,Chen:2020ody} RG resummation (RGR)~\cite{Su:2022fiu} and leading renormalon resummation (LRR)~\cite{Zhang:2023bxs}, within the window $a<z\ll1/\Lambda_{\rm QCD}$. Therefore, perturbative uncertainties in $C_0$ are propagated into the systematic uncertainty of $m_0$.

\begin{figure*}[htbp]
    \centering
    \includegraphics[width=0.8\linewidth]{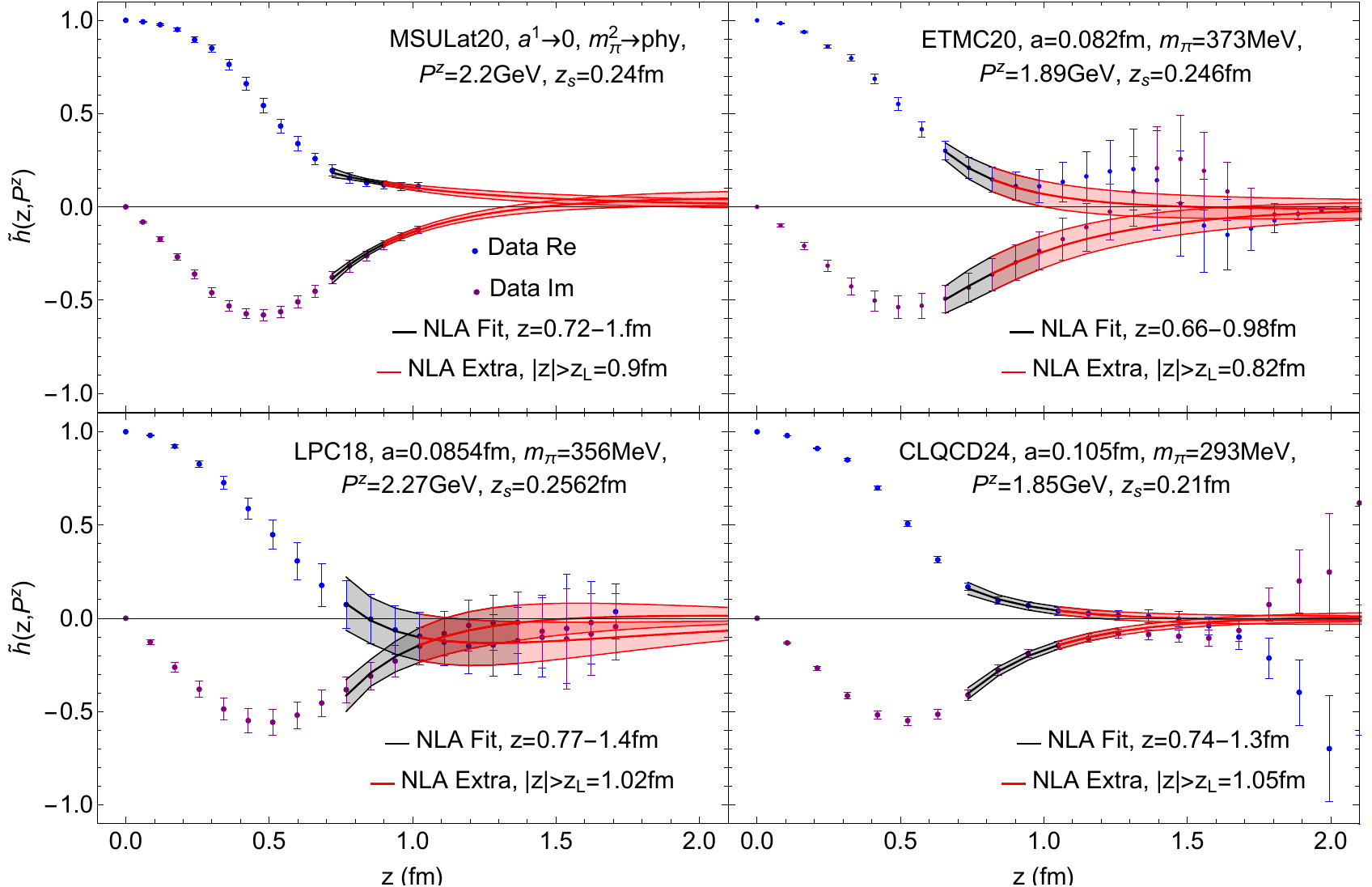}
    \caption{Large distance asymptotic analysis. The blue and purple points denote the real and imaginary parts of the lattice data, respectively. The black bands cover the fit regions in the asymptotic analysis, while the red bands show the extrapolated large-distance behavior. }
    \label{fig:fourier}
\end{figure*}

Lattice artifacts, such as excited-state contaminations, unphysical pion masses, and discretization effects, could distort the global behavior of the lattice PDF, causing a larger first moment $\langle x \rangle$ than the phenomenological value~\cite{Mondal:2020ela,Alexandrou:2020qtt,Djukanovic:2024krw}. To mitigate these effects, we decide to enforce the first moment of the PDF to the state-of-the-art lattice-determined physical value from Ref.~\cite{Djukanovic:2024krw}, which is consistent with other recent lattice calculations~\cite{Mondal:2021oot,Alexandrou:2026oks} and phenomenological results~\cite{Harland-Lang:2014zoa,Hou:2019efy}. In coordinate space, it is the slope of the imaginary part of the correlator near $z=0$, namely the coefficient of $i z P^z$ according to the operator product expansion (OPE). Since the correction factor $1+i\delta y zP^z$ at large $z$ has an unrealistically large
imaginary part, we unitarize it 
to an exponential. We thus use the following ansatz, 
\begin{align}\label{eq:CalM}
\tilde{h}(z,P^z) = \tilde{h}^{R}(z,P^z) e^{i z P^z \delta y}  \ ,
\end{align}
to remove a large part of the lattice artifacts. The parameter $\delta y$ is determined by fitting the imaginary part of $\tilde{h}(z,P^z)$ to the short distance OPE~\cite{Radyushkin:2017cyf,Braun:2007wv,Izubuchi:2018srq},
\begin{align}\label{eq:SDEim}
- i z P^z \, \langle x \rangle(\mu) \, \frac{C_{1}(z^2 \mu^2)}{C_{0}(z^2 \mu^2)} + ...  \quad \text{for }  a<z\ll 1/\Lambda_{\rm QCD} \ ,
\end{align}
where the first moment is fixed to the Mainz24 value~\cite{Djukanovic:2024krw}, $\langle x \rangle(\mu=2\, {\rm GeV})=0.153(17)$.
The Wilson coefficients $C_{i}(z^2 \mu^2)$~\cite{Izubuchi:2018srq} are evaluated at NNLO+RGR+LRR accuracy. We find $\delta y = 0.104(16)$, $0.094(16)$, $0.068(16)$, and $0.085(17)$ for the four datasets, respectively, where the uncertainty originates from the first moment. The coordinate-space results $\tilde{h}(z,P^z)$ are shown as the blue and purple points in Fig.~\ref{fig:fourier}.


The next step is to Fourier transform (FT) the spatial correlator into momentum space. Formally, the FT requires the correlator over the entire $z$ range, including the large-distance region (e.g. $\gtrsim1$ fm). Because of confinement, the correlator decays exponentially at large distances, so that the long-distance tail contributes only weakly to the FT. In practice, however, lattice data at large $z$ suffer from exponentially deteriorating signal-to-noise ratios and cannot be directly used. This issue has been investigated through asymptotic analyses in Refs.~\cite{Ji:2020brr,Gao:2021dbh,Chen:2025cxr}. A recent work~\cite{Ji:2026vir} derives the large-distance asymptotic forms for various quasi-correlators, which serve as physical constraints on the FT. For the unpolarized $u-d$ proton quasi-PDF matrix element, the leading asymptotic (LA) and next-to-leading asymptotic (NLA) forms in the $z\to\infty$ expansion are
\begin{align}
\tilde{h}^{\rm LA}\left(z,P^z\right) = 
A e^{i \phi \, {\rm sign}(z)}  e^{-\Lambda |z|}  \ ,
\end{align}
\begin{align}
\tilde{h}^{\rm NLA}\left(z,P^z\right) = 
\left[A e^{i \phi \, {\rm sign}(z)} 
+ \frac{ A' e^{i \phi' \, {\rm sign}(z)} }{|z|} \right] e^{-\Lambda |z|}  \ ,
\end{align}
where $A$, $\phi$, $\Lambda$, $A'$, and $\phi'$ are real fit parameters. The mass gap $\Lambda$ corresponds to the binding energy of an intermediate state within the quasi-PDF correlator, which is a vector heavy-light meson. Its value is estimated as $400-600$ MeV, and we impose a loose constraint on it during the fit to ensure stability. We fit the asymptotic form $\tilde{h}_{\rm asym}(z,P^z)$ ($\tilde{h}_{\rm asym}$ denotes either $\tilde{h}^{\rm LA}$ or $\tilde{h}^{\rm NLA}$) to the lattice data $\tilde{h}(z,P^z)$ at distances larger than $\sim 1/\Lambda$ (e.g. $0.7$--$1.4$ fm), with representative NLA fits shown as the black bands in Fig.~\ref{fig:fourier}. Uncertainties in the fit parameters will be propagated into the statistical uncertainty. The differences between the LA and NLA analyses, which are at the few-percent level, will be included in the systematic uncertainty.   
The quasi-PDF is then obtained through the FT,
\begin{align}\label{eq:FT}
\tilde{f}\left(y,P^z\right) = &
P^z \int_{-\infty}^{+\infty} \frac{d z}{2\pi} e^{i z P^z y} \left[ \tilde{h}(z,P^z) \theta(z_L-|z|) \right. \nonumber\\
&\left. + \tilde{h}_{\rm asym}(z,P^z) \theta(|z|-z_L) \right] \ ,
\end{align}
where the switching point $z_L$ is chosen around $0.8$--$1$ fm.

\begin{figure*}[htbp]
    \centering
    \includegraphics[width=0.8\linewidth]{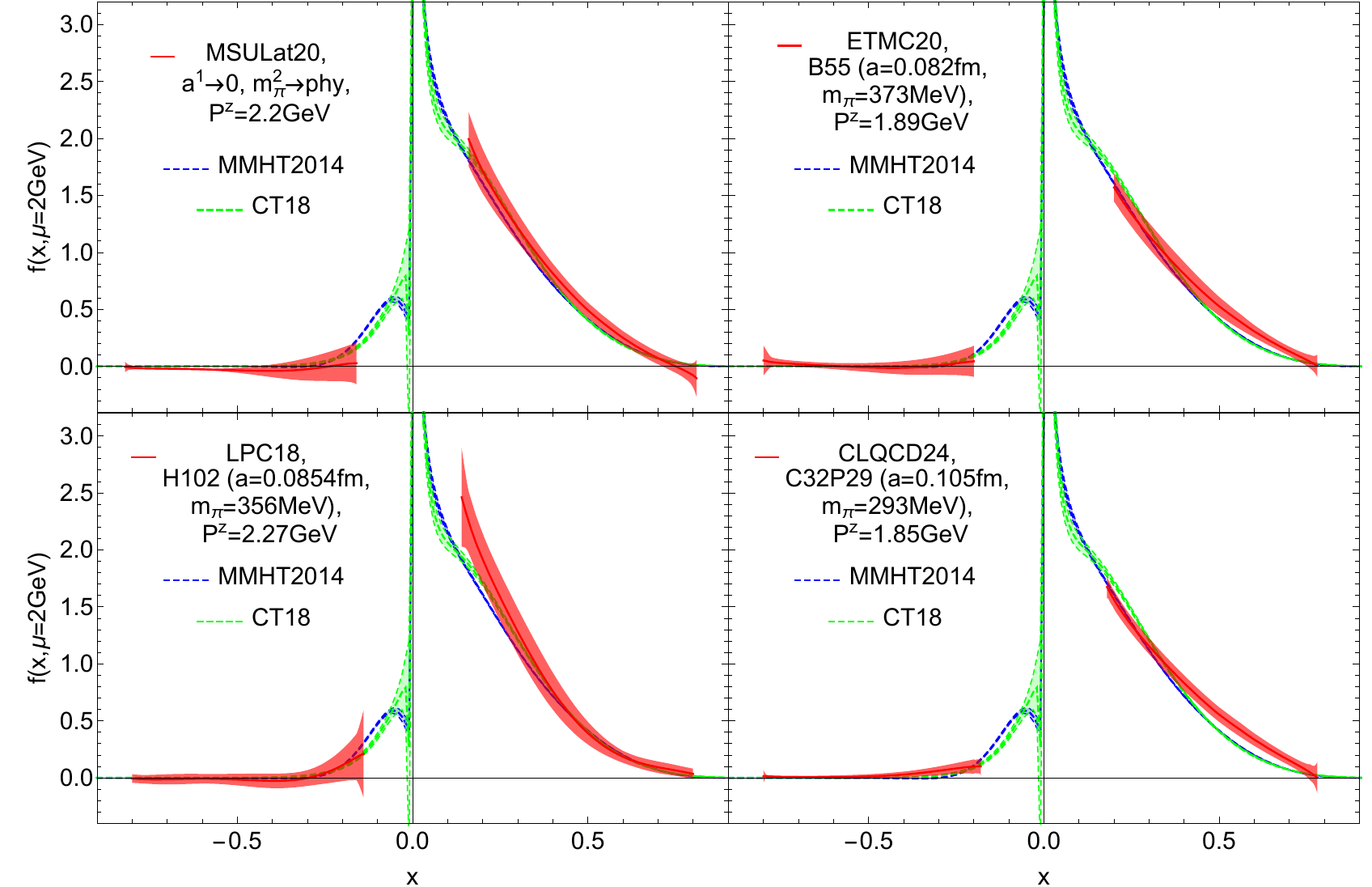}
    \caption{The proton isovector unpolarized PDFs. The red bands denote the LaMET results in the region $\Lambda_{\rm QCD}/2P^z \lesssim x \lesssim 1-\Lambda_{\rm QCD}/2P^z$, where the perturbative matching is reliable. The blue and green bands correspond to the phenomenological PDFs extracted from global analyses by MMHT2014~\cite{Harland-Lang:2014zoa} and CT18~\cite{Hou:2019efy}, respectively.  }
    \label{fig:lcPDF}
\end{figure*}

The PDF $f$ and quasi-PDF $\tilde{f}$ share the same IR physics but differ in UV physics~\cite{Xiong:2013bka,Ma:2014jla,Ma:2017pxb,Izubuchi:2018srq,Wang:2019tgg,Ji:2020ect,Holligan:2025baj}. They are connected through a LaMET expansion~\cite{Ji:2024oka},
\begin{align}\label{eq:matching}
f(x,\mu) = &\int_{-\infty}^{+\infty} \frac{d y}{|y|} C\left(\frac{x}{y},\frac{\mu}{y P^z}\right) \tilde{f}(y,P^z) \nonumber\\
&+ O\left(\frac{\Lambda_{\rm QCD}^2}{x^2 P_z^2}, \frac{\Lambda_{\rm QCD}^2}{(1-x)^2 P_z^2}\right) \ ,
\end{align}
where the perturbative matching kernel $C\left(x/y,\mu/y P^z\right)$ is at NNLO accuracy~\cite{Li:2020xml,Chen:2020ody}, supplemented by DGLAP log resummation (RGR)~\cite{Su:2022fiu}, leading-renormalon resummation (LRR)~\cite{Zhang:2023bxs}, and threshold log resummation (TR)~\cite{Ji:2023pba,Liu:2023onm,Ji:2024hit}. The RGR and TR primarily control the precision at small and large $x$, respectively. The LRR improves perturbative convergence. Moreover, the renormalon-regularization scheme of the matching kernel is chosen consistently with the constant $m_0$ in $Z_R$ of Eq.~(\ref{eq:HybridM}), ensuring the cancellation of the linear power correction~\cite{Zhang:2023bxs}. Systematics arising from omitted higher-order perturbative corrections are estimated by varying the physical scales that enter the resummation procedures.

\textit{Results:}\quad The LaMET results for the proton $u-d$ unpolarized PDF are shown as the red bands in Fig.~\ref{fig:lcPDF}. For ETMC20, LPC18, and CLQCD24 datasets, the uncertainties involve statistical fluctuations, the renormalization parameter $m_0$, first-moment calibration, Fourier transform, and perturbative matching. Particularly, the uncertainty associated with finite hadron momentum $P^z$ is estimated through omitted higher-order terms in the perturbative matching rather than through higher-twist corrections. For the MSULat20 dataset, an additional systematic uncertainty from the continuum and physical pion-mass extrapolations is estimated from the spread among various ansätze in Fig.~4 of Ref.~\cite{Lin:2020fsj}. The detailed explanations of the uncertainty estimate and propagation are included in the supplementary material.

After applying a common state-of-the-art framework to all four lattice datasets, the spread among the resulting PDFs is reduced compared with the original analyses, as shown in the supplementary material. Since phenomenological PDFs~\cite{ATLAS:2021vod,NNPDF:2021njg,Bailey:2020ooq,Hou:2019efy,Alekhin:2017kpj,H1:2015ubc,Jimenez-Delgado:2014twa} are nearly identical in the moderate-$x$ region, only two representative results are shown in Fig.~\ref{fig:lcPDF} for comparison. The continuum- and physical-mass-extrapolated MSULat20 result agrees with the phenomenological PDFs within approximately $1\sigma$ in the accessible $x$-region. The remaining datasets exhibit broadly compatible $x$-dependence, with small deviations that may arise from residual discretization effects or unphysical pion masses.
Additional lattice data will help to confirm this
expectation. According to our numerical tests, the MSULat20 results for each ensemble before the extrapolations are similar to those of ETMC20 and CLQCD24. The ratios of the LaMET and global fit PDFs are shown in the supplementary material to further illustrate the above observations. 

\textit{Summary:}\quad We present a LaMET calculation of the proton isovector unpolarized PDF using existing lattice QCD datasets, along with theoretical precision-control techniques and an empirical method to remove a large part of
lattice artifacts. The disagreement among earlier lattice PDF calculations appears to be largely attributable to analysis methodology and lattice artifacts. The resulting PDF for the physical-limit-extrapolated dataset, MSULat20, agrees with phenomenological determinations, while other datasets show broad alignment, supporting LaMET's potential to achieve quantitatively reliable determinations of the $x$-dependence of parton distributions. In particular, hadron momenta of order $P^z\sim 2$ GeV may already be sufficient for precision studies when combined with the latest analysis techniques. This work is an important step towards precision LaMET calculations, which will offer essential inputs for high-energy experiments at the Large Hadron Collider (LHC), Jefferson Laboratory (JLab), and Electron-Ion Colliders (EICs).

\section*{Acknowledgement}
We thank ETMC, LPC, and CLQCD for sharing lattice data. We thank Jinchen He, Jinghong Yang, Jun Hua, Yong Zhao, Rui Zhang, Andreas Schäfer, Jiunn-Wei Chen, and Huey-Wen Lin for useful discussions and comments. Y. S. is partially supported by the Quark-Gluon Tomography (QGT) collaboration, which is supported by the U.S. Department of Energy (DOE) topical collaboration program (DE-SC0023646). X. Ji is partially supported by a grant from National Science Foundation of China, No. 12635004 and Thomas and Linda Lau Family Foundation.  

\bibliography{bibliography}

\clearpage

\onecolumngrid

\section*{Supplementary material}

This supplementary material provides evidence supporting the methods and arguments in the main text, including empirical mitigation of lattice artifacts, comparisons among lattice groups, comparisons between lattice and phenomenology, and uncertainty estimates. 

\begin{figure}
    \centering
    \includegraphics[width=0.5\linewidth]{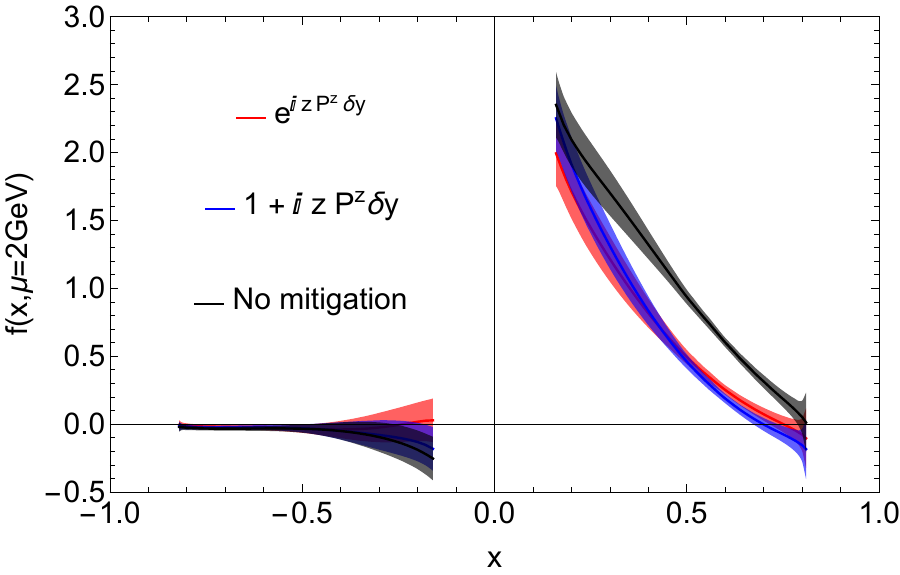}
    \caption{Proton's unpolarized isovector PDFs calculated using different forms to mitigate lattice artifacts. The dataset is the MSULat20~\cite{Lin:2020fsj} extrapolated to the physical limit. The red band is related to the exponential mitigation form adopted in the main context. The blue band corresponds to the linear correction form. The black band represents the result without mitigation. }
    \label{fig:f_MSU20_compareforms}
\end{figure}
Fig.~\ref{fig:f_MSU20_compareforms} tests whether the PDF results are sensitive to the choice of correction forms to mitigate lattice artifacts. The red band shows the PDF result related to the exponential form in Eq.~(\ref{eq:CalM}) of the main text, $e^{i z P^z \delta y}$. The blue band corresponds to the linear form $1 + i z P^z \delta y$. The two results are consistent within uncertainties, indicating that the resulting PDFs are not strongly sensitive to the two correction forms considered here. For reference, the black band does not mitigate the lattice artifacts and deviates from the others, indicating the importance of removing genuine lattice artifacts.  

\begin{figure}
    \centering
    \includegraphics[width=0.48\linewidth]{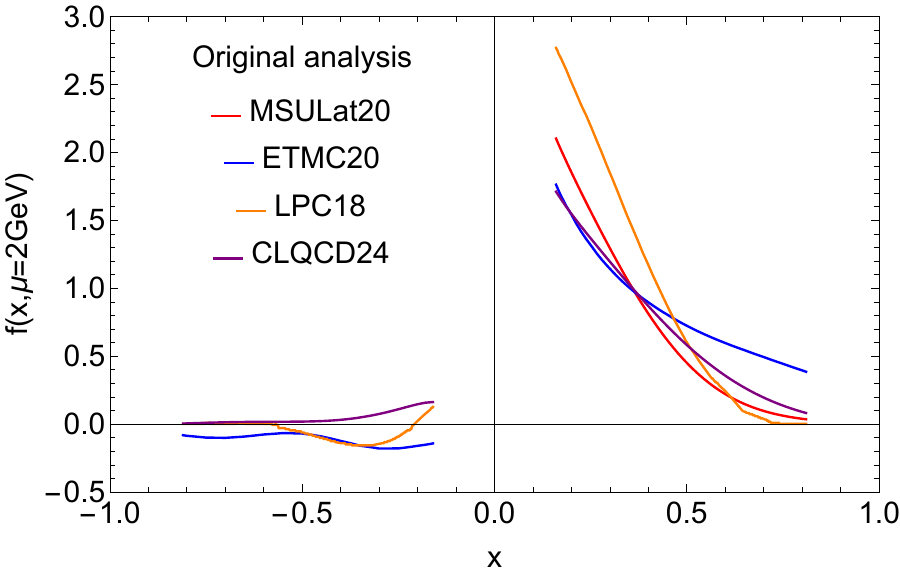}
    \includegraphics[width=0.48\linewidth]{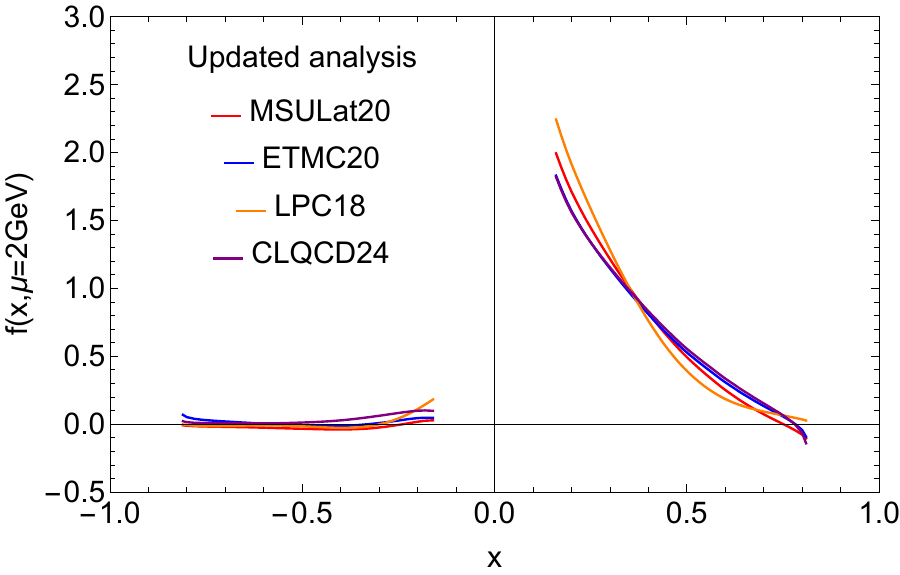}
    \caption{Comparison between the analysis in the original literature~\cite{Lin:2020fsj,Alexandrou:2020qtt,LatticeParton:2018gjr,Chen:2024rgi} (left panel) and the reanalysis in this work (right panel). Only the central values are displayed. The spread across different lattice groups is reduced under the updated analysis.}
    \label{fig:f_before_after}
\end{figure}
Fig.~\ref{fig:f_before_after} compares the resulting lattice PDFs before and after the reanalysis. The left panel corresponds to the original analysis in Refs.~\cite{Lin:2020fsj,Alexandrou:2020qtt,LatticeParton:2018gjr,Chen:2024rgi}. The red, blue, and orange curves are digitized from the relevant plots in Refs.~\cite{Lin:2020fsj,Alexandrou:2020qtt,LatticeParton:2018gjr}, respectively. Since Ref.~\cite{Chen:2024rgi} does not directly display the $u-d$ combination, we calculate the purple curve using the data and methods in Ref.~\cite{Chen:2024rgi}. The right panel shows the updated analysis in this work, for which the spread among the lattice datasets is substantially reduced.   

\begin{figure}
    \centering
    \includegraphics[width=0.8\linewidth]{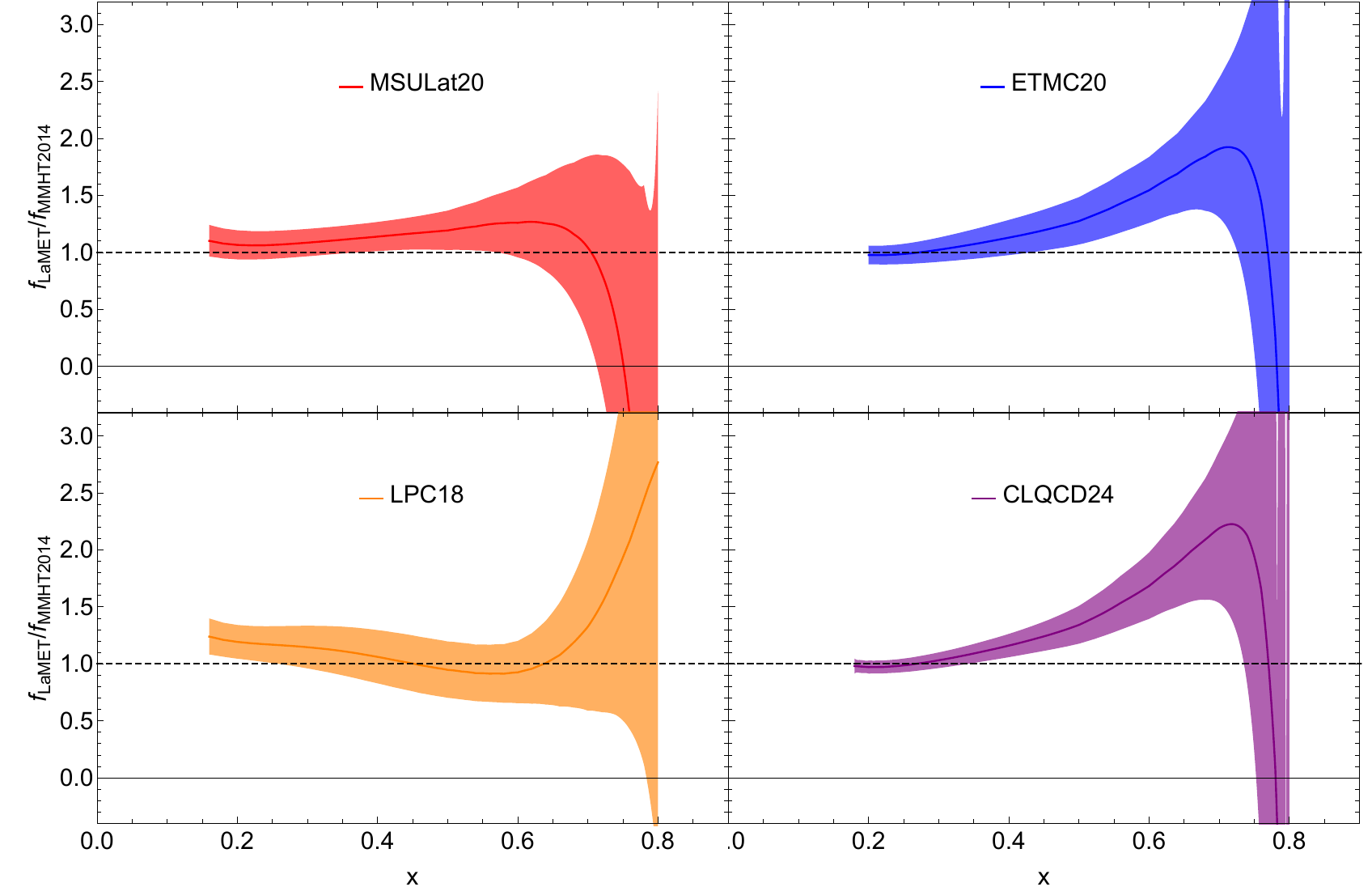}
    \caption{The ratios between PDFs from LaMET and MMHT2014~\cite{Harland-Lang:2014zoa}. }
    \label{fig:f_ratio}
\end{figure}
To better compare LaMET and global fit PDFs, we plot their ratios in Fig.~\ref{fig:f_ratio}. The LaMET result from the physical-limit dataset, MSULat20, shows $\sim 1 \sigma$ agreement with the global fit PDF. The other datasets are broadly compatible with the global fit, with residual deviations that may arise from discretization or unphysical pion-mass effects. 

\begin{figure}
    \centering
    \includegraphics[width=0.8\linewidth]{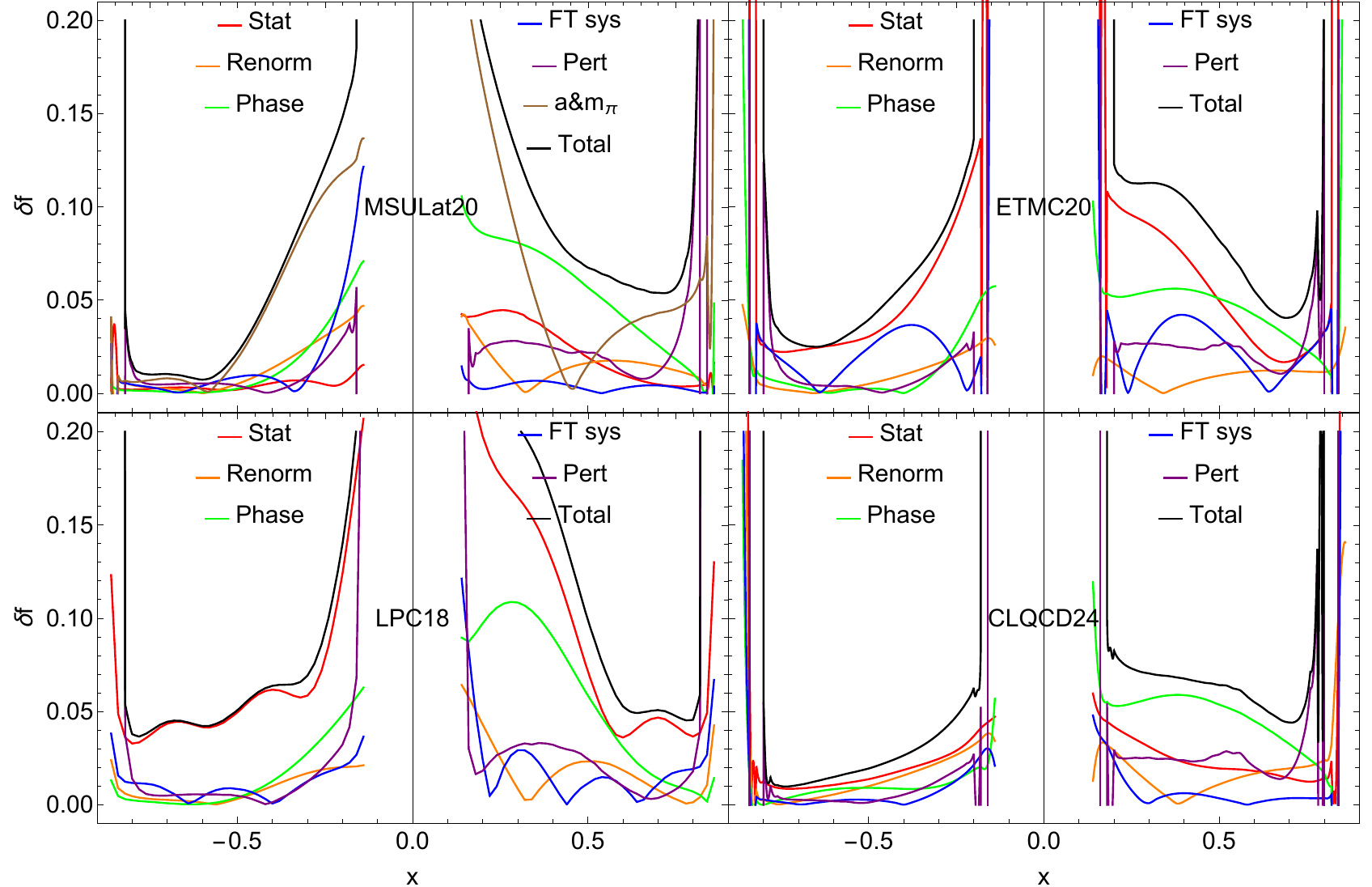}
    \caption{Uncertainties for proton's unpolarized isovector PDFs in Fig.~\ref{fig:lcPDF}. The red curve is the statistical uncertainty. The orange curve arises from the uncertainty of $m_0$ in the renormalization. The green curve is the uncertainty from the phase rotation parameter $\delta y$. The blue curve is the systematic uncertainty of the Fourier transform estimated as the difference between LA and NLA. The purple curve denotes the uncertainty of perturbative matching evaluated through the variations of physical scales in the RG evolution studied in Ref.~\cite{Ji:2024hit}. The brown curve is the systematics from the continuum and physical pion mass extrapolations, estimated using the deviation among various ansätze in Fig.~4 of Ref.~\cite{Lin:2020fsj}. The black curve is the total uncertainty obtained by treating the individual contributions as uncorrelated and adding them in quadrature. }
    \label{fig:f_err_all}
\end{figure}
Fig.~\ref{fig:f_err_all} shows the detailed uncertainty estimates. The red curves correspond to statistical uncertainties $\sigma_{\rm stat}$, with ETMC20, LPC18, and CLQCD24 calculated using bootstraps, which automatically take into account data at $|z|<z_L$ and the extrapolated form at $|z|>z_L$, as well as their correlations. For MSULat20, we combine in quadrature the statistical uncertainties digitized from Fig.~5 in Ref.~\cite{Lin:2020fsj} with those propagated from the fitted large-distance forms. We find that the former source of statistical uncertainty is much larger than the latter, e.g., 0.019 vs. 0.003 at $x=0.5$, indicating that the uncorrelated combination of the two sources~\footnote{The correlation within each source is considered.} will not deviate much from the correlated one. The orange curves $\sigma_{\rm renorm}$ arise from the uncertainty of $m_0$ in the renormalization. Since $m_0$ is determined by fitting the perturbative Wilson coefficients, the perturbative uncertainties are propagated into $m_0$, which eventually affects the momentum-space PDFs. The green curves $\sigma_{\rm phase}$ come from the uncertainties of $\delta y$ in the mitigation form, which originate from the first moment~\cite{Djukanovic:2024krw}, $\langle x \rangle(\mu=2\, {\rm GeV})=0.153(17)$. The blue curves $\sigma_{\rm FT \, sys}$ are the systematic uncertainties of the Fourier transform, estimated as the difference between LA and NLA, following the method in Ref.~\cite{Ji:2026vir}. The purple curves $\sigma_{\rm pert}$ originate from perturbative matching uncertainties, estimated by varying the physical scales in the RG evolution following Ref.~\cite{Ji:2024hit}. The brown curve $\sigma_{a\& m_{\pi}}$ for MSULat20 denotes the systematics from the physical extrapolations, estimated using the deviation among various ansatze in Fig.~4 of Ref.~\cite{Lin:2020fsj}. When we calculate the uncertainty for one step, the parameters at other steps are set to their central values. The total uncertainty $\sigma_{\rm tot}$ is obtained by treating these contributions as uncorrelated and adding them in quadrature,
\begin{align}
\sigma_{\rm tot} = \sqrt{ \sigma_{\rm stat}^2 + \sigma_{\rm renorm}^2 + \sigma_{\rm phase}^2 + \sigma_{\rm FT \, sys}^2 + \sigma_{\rm pert}^2 + \sigma_{a\& m_{\pi}}^2} \ ,
\end{align}
where $\sigma_{a\& m_{\pi}}$ is considered only for MSULat20 and not included for other datasets.

\end{document}